\documentstyle[12pt]{article}

\newcommand{\onehalf}{\mbox{$\frac{1}{2}$}}
\newcommand{\no}[1]{: \! #1 \! :}
\newcommand{\nsp}{\!\!\!\!\!}
\newcommand{\vJ}{\mbox{\boldmath $J$}}
\newcommand{\vS}{\mbox{\boldmath $S$}}
\newcommand{\vphi}{\mbox{\boldmath $\phi$}}
\newcommand{\Z}{{\bf Z}}
\newcommand{\vone}{{\bf 1}}
\newcommand{\phdagger}{\mathop{\phantom{\dagger}}}
\newcommand{\psiop}[1]{\psi^{\phdagger}_{#1}}
\newcommand{\psidop}[1]{\psi^{\dagger}_{#1}}

\newcommand{\bsigma}[1]{\mbox{\boldmath $\sigma$}^{\phdagger}_{#1}}

\begin{document}
\sloppy
\renewcommand{\thefootnote}{\fnsymbol{footnote}}
\noindent{\large\bf Interacting Electrons and Localized Spins: \\
Exact Results from Conformal Field Theory\footnote{{\tt cond-mat/9506098},
to appear in the Proceedings of the 1995 Schladming Winter School on
Low-Dimensional Models in Statistical Physics and Quantum Field Theory.}}\\
\\
Per Fr\"ojdh$^1$ and Henrik Johannesson$^2$\\ \\
{\small\em $^1$Department of Physics, University of Washington,
P.O. Box 351560, Seattle, WA 98195, USA}\\
{\small\em $^2$Institute of Theoretical Physics, Chalmers University
of Technology and G\"oteborg University, S-412 96 G\"oteborg, Sweden}\\

\subsection*{Introduction}

Much of traditional condensed matter physics falls under the Fermi
liquid paradigm. The notion goes back to Landau and implies
that a liquid of fermions (such as $^3$He or conduction electrons in a
metal) can be treated as a system of essentially free particles \cite{AGD}.
Certain conditions apply: symmetries must remain unbroken and the
energies probed should be ``small.'' But with these provisos the only price
to pay for removing the interactions is to keep track on how certain
parameters (quasiparticle masses, lifetimes, etc.) renormalize.
The Fermi-liquid picture has proven enormously successful and explains how
we can get away with single-particle quantum mechanics when doing
elementary solid state physics.
Fortunately --- to keep us busy! --- experimentalists have found a number of
lab systems that violate the standard Fermi liquid picture.
(An outstanding example is the metallic phase of the high-$T_c$
superconductors.) The design of concepts and methods to handle this
intriguing and growing class of problems --- often nicknamed
``strongly correlated systems'' --- is a major challenge for the theorist.
By the very nature of the problem one here confronts matter
nonperturbative. Lacking a universal paradigm, the most sensible thing
to do is to practice on simple models of non-Fermi liquids,
work out the consequences, compare with experiments, build up intuition,
and collect that ``critical mass'' of knowledge necessary for
the emergence of an effective, unified approach to correlated systems.

Two particularly clean realizations of non-Fermi liquids are the
Luttinger liquid and (certain generalized versions of) the Kondo
effect. ``Luttinger liquid'' is a code name for the low-energy,
long-wavelength physics of interacting electrons in one dimension
(1D) \cite{Schulz}. The topology of the 1D Fermi surface
for free electrons (it consists of two distinct points!) yields
dramatic effects when interactions are included: The single-particle
poles of the propagators get wiped out and are replaced by branch cuts.
The resulting spectral density develops a two-peak structure,
with a broad band in between: A single electron added to the system
``decays'' into two distinct collective excitations, one carrying the
charge, the other the spin. Remarkably, the two excitations propagate
with different velocities, leading to a spatial spin-charge separation.
The Kondo effect is a different story \cite{Hewson}. Here one
considers the spin exchange interaction between a localized magnetic
impurity and a band of free conduction electrons in a metal.
By symmetry, the problem is identical to that of 1D free fermions coupled
to a local spin.
Surprisingly, this mundane setup leads to quite spectacular physics.
As one scales to low temperatures and large distances, the
electron-impurity coupling flows to a strong-coupling fixed point where
electrons are effectively interacting with each other {\em via the
impurity}. In certain versions of the problem, where the impurity couples
to several degenerate bands of conduction electrons (multi-channel Kondo
effect), the induced interaction among electrons cannot be
``renormalized away,'' and one finds a new kind of low-temperature
critical behavior. Yet another breakdown of the simple Fermi-liquid
picture!

The theories of Luttinger liquids and the multi-channel Kondo
effect are important landmarks in the physics of correlated systems. Their
construction did not come easy, though, and have required decades of
shrewd work by many people, exploiting some of the most powerful
tools of theoretical physics: Bosonization, the Renormalization Group,
Bethe's {\em Ansatz}, and most recently Conformal Field Theory (CFT).

\subsection*{Boundary Conformal Field Theory}

The use of CFT to study the Kondo effect was pioneered by Affleck and
Ludwig in a series of remarkable papers \cite{Affleck}. The basic idea
goes back to Nozi\`{e}res \cite{Nozieres}. In the simplest case --- with a
spin-$\onehalf$ impurity coupled antiferromagnetically to a single channel
of conduction electrons --- Nozi\`{e}res showed
that the impurity may be traded for a boundary condition: At the
strong-coupling fixed point the impurity traps an electron into
a bound singlet state, with the remaining electrons being excluded
from entering the impurity site. In other words, the impurity disappears
from the problem and is replaced by a boundary condition on the free
electron wave functions.
Affleck and Ludwig argued that all quantum impurity problems essentially
work this way, and showed how to take advantage of CFT once the problem
is reformulated in terms of a new boundary condition.

By introducing a boundary in CFT, say at $x=0$, one imposes constraints
on any field $\varphi(t,x)$ \cite{Cardy1}. The left and right moving pieces get
identified
via analytic continuation beyond the boundary, such that
\begin{equation}
\varphi(t,x) = \varphi_L(t,x)\varphi_R(t,x) \ \rightarrow \
\varphi_L(t,x)\varphi_L(t,-x) .
\end{equation}
Hence, the boundary has effectively turned the field nonlocal. As a
consequence, the one-point function in the presence of a boundary
develops a profile given by the corresponding two-point function in the bulk,
\begin{equation}
\langle \varphi(t,x) \rangle \ \rightarrow \
\langle \varphi_L(t+x) \varphi_L(t-x) \rangle \
\sim \ x^{-\Delta_{bulk}} ,
\end{equation}
with $\Delta_{bulk}$ the ordinary scaling dimension of $\varphi$. Correlation
functions are also altered. In particular, the autocorrelation function
for $\varphi$ close to the boundary is governed by a boundary-condition
dependent exponent $\Delta$ called the {\em boundary scaling dimension}
of $\varphi$:
\begin{equation}
\langle \varphi(t,x) \varphi(0,x) \rangle - \langle \varphi(t,x) \rangle
\langle \varphi(0,x) \rangle \ \sim \ t^{-2\Delta} \ , \ \ \
\ \ \ |t| >> |x|. \label{auto}
\end{equation}

This sets the strategy for treating quantum impurity problems: Identify first
the particular boundary condition that plays the role of the impurity
interaction. (In most cases this is a highly nontrivial task.
The CFT scheme gives some guidance, though, and it has turned out that
{\em conformal fusion} often corresponds to such a change of boundary
condition.) Given that the new  boundary condition is indeed
scale invariant (together with the original bulk theory), one can then
use the machinery of CFT to extract the corresponding boundary
scaling dimensions. As these determine the asymptotic autocorrelation
functions, the finite-temperature properties due to the presence of the
boundary ({\em alias} the impurity!) are easily accessed from standard
finite-size scaling by treating (Euclidean) time as an inverse temperature.

\subsection*{Kondo effect in a Luttinger Liquid}

What happens if one puts a magnetic
impurity into a Luttinger liquid (LL)? As for the Kondo effect in a Fermi
liquid we expect that the impurity also here will induce effective
interactions among the electrons. But in a LL, the electrons are already
strongly correlated by the mutual Coulomb interactions. Then, what
happens? Does the interplay between ``induced'' and ``direct'' correlations
lead to novel effects? Or do we recover the same Kondo physics as in a
Fermi liquid? This is not an easy problem, but again, CFT can be put to
productive use \cite{FJprl}.

The ``standard model'' for low-energy electrons
in 1D is the Tomonaga-Luttinger (TL)
Hamiltonian \cite{Schulz}
\begin{eqnarray}
{\cal H}_{TL}  =  \frac{1}{2 \pi} \int dx \biggl\{ & \nsp v_F
\nsp & \biggl[ \no{ \psidop{L,\sigma}(x) i \frac{d}{dx} \psiop{L,\sigma}(x) }
- \no{ \psidop{R,\sigma}(x) i \frac{d}{dx} \psiop{R,\sigma}(x) } \biggr]
 \nonumber \\
 & \nsp + \nsp & \frac{g}{2} \sum_{k, l = L, R}^{\phantom{x}}
  \no{ \psidop{k,\sigma}(x) \psiop{k,\sigma}(x) }
\no{ \psidop{l, -\sigma}(x) \psiop{l, -\sigma}(x) } \nonumber \\
 & \nsp + \nsp & \, g \, \no{ \psidop{R,\sigma}(x) \psiop{L,\sigma}(x)
\psidop{L,-\sigma}(x) \psiop{R,-\sigma}(x) } \biggr\} , \ \ \ \ g >
0. \label{TL}
\end{eqnarray}
Here $\psi_{L/R,\sigma}(x)$ are the left/right moving components
of the electron field $\Psi_{\sigma}(x)$, expanded about the
Fermi points $\pm k_F$, and we implicitly sum over repeated indices for
spin $\sigma = \uparrow, \downarrow$. The first term in (\ref{TL})
is that of free relativistic fermions, while the second
and third terms describe forward and backward electron-electron
scattering, respectively. Normal ordering is carried out w.r.t. the
filled Dirac sea.

The TL Hamiltonian is conveniently written on diagonal Sugawara
form, using the charge and spin currents
\begin{eqnarray}
j_{L/R}(x) & = &
\cosh\theta \no{\psidop{L/R,\sigma}(x) \psiop{L/R,\sigma}(x) } +
\sinh\theta \no{\psidop{R/L,\sigma}(x) \psiop{R/L,\sigma}(x) }, \\
\vJ_{L/R}(x) & = &
\no{\psidop{L/R,\sigma}(x) \mbox{$\frac{1}{2}$} \psiop{L/R,\mu}(x)},
\label{vJcurrent}
\end{eqnarray}
with $\tanh2\theta = g/(v_F+g)$. Dropping a marginally irrelevant term
$-(g/\pi) \vJ_L \cdot \vJ_R$, one obtains the critical bulk Hamiltonian
\begin{equation}
\label{Sugawara}
{\cal H}^{*}_{TL} = \frac{1}{2\pi} \int_{-\ell}^{\ell} dx \sum_{i =
1,2} \left\{
\frac{v_c}{4} \no{j^{i}_L(x)j^{i}_L(x)}
+ \frac{v_s}{3} \no{\vJ^{i}_L(x) \cdot \vJ^{i}_L(x)} \right\},
\end{equation}
where we have confined the system to $x \in [-\ell,\ell ]$ and replaced
the right-moving currents with a second channel of left-moving currents.
The last step involves folding the interval in half to $[0,\ell ]$ and
analytically continue the left-handed currents back to the full interval.
Note that the spin and charge separation in (\ref{Sugawara}) yields two
dynamically independent theories, each Lorentz invariant with a
characteristic velocity, $v_c = v_F (1 + 2g/v_F)^{\frac{1}{2}}$ and
$v_s = v_F - g$. The charge and spin currents satisfy the $U(1)$ and
(level-1) $SU(2)_1$ Kac-Moody algebras, respectively, and it follows
that ${\cal H}^{*}_{TL}$ is invariant under the chiral symmetry
$U(1) \times U(1) \times SU(2)_{1} \times SU(2)_{1}$.

We now couple a localized spin-\onehalf\ impurity $\vS$ to the electrons.
As a warm-up
let us first consider the simple case of only {\em forward scattering}
off the impurity:
\begin{equation}
{\cal H}_{F} = \sum_{k=L,R} \lambda \no{ \psidop{k,\sigma}(0)
\mbox{$\frac{1}{2}$} \bsigma{\sigma \mu} \psiop{k,\mu}(0) }
\cdot \vS .
\label{F}
\end{equation}
Using the spin currents in (\ref{vJcurrent}), and replacing the
right-handed one by a second left-handed current, Eq. (\ref{F}) turns into
\begin{equation}
{\cal H}_{F} = \lambda_F [ \vJ_L^1(0) + \vJ_L^2(0) ] \cdot \vS.
\label{newFORWARD}
\end{equation}
As the two currents are coupled via $\vS$, ${\cal H}_{F}$
breaks the $SU(2)_1 \times SU(2)_1$ symmetry of ${\cal H}^*_{TL}$ down to
the diagonal level-2 subalgebra $SU(2)_2$ spanned by
$\vJ (x) = \vJ_L^1(x) + \vJ_L^2(x)$. To adopt to this fact we use the
coset construction \cite{Goddard} to write the spin part
of the Hamiltonian as a sum of an $SU(2)_2$ Sugawara Hamiltonian
and a free Majorana fermion ($\Leftrightarrow$ 2D Ising model).
We can then absorb ${\cal H}_F$ into ${\cal H}^{*}_{TL}$
by the canonical transformation
$\vJ (x) \rightarrow \vJ '(x) \equiv \vJ (x) + \vS \delta (x) $,
$\vJ '(x)$ being the spin current of electrons {\em and} impurity.
The impurity thus disappears from the Hamiltonian and is replaced by a
nontrivial boundary condition on ${\cal H}^{*}_{TL}$.

In CFT, a boundary condition is equivalent to a selection rule for
quantum numbers of a conformal embedding. In our case, the conformal
embedding is $U(1) \times U(1) \times SU(2)_2 \times \mbox{{\em Ising}}$,
with quantum numbers $Q,\Delta Q \in \Z$ (sum and difference of net charge
in the two channels), $j = 0, \onehalf, 1$ (spin of primary states),
and $\phi = \vone, \sigma, \epsilon$ ({\em Ising} primary fields).
Without the impurity, i.e. with
a trivial boundary condition at $x=0$, the eigenstates of ${\cal
H}^{*}_{TL}$ organize into products of four conformal towers, labeled by
these quantum numbers. The allowed combinations can be described by a
selection rule that reproduces the spectrum of interacting fermions.
{\em With} the impurity (nontrivial boundary condition), the spin current gets
redefined as that of electrons {\em and} impurity: Effectively, a
spin-\onehalf\ degree of freedom is added to the $SU(2)_2$ towers, which
get shifted according to the CFT fusion rule $j=0 \rightarrow \onehalf,
\onehalf \rightarrow 0$ or $1, 1 \rightarrow \onehalf$.
A new spectrum emerges, with the $U(1) \times U(1)$, $SU(2)_2$, and
Ising conformal towers combined differently according to a new selection
rule. It describes {\em interacting fermions + impurity} and yields
immediate information about the  boundary scaling dimensions
(which are really what we are after!). This
follows from a well-known result by Cardy \cite{Cardy2}: The eigenstates
of a Hamiltonian ${\cal H}$ on a finite interval are in 1-1 correspondence
to the boundary scaling dimensions of the same theory boosted to
Euclidean space-time (with the time axis defining the boundary).

Having thus obtained the spectrum of boundary scaling dimensions, we
identify the leading correction-to-scaling boundary operator
$(LCBO)$ ${\cal O}$
that respects all symmetries of the problem. This operator is added to
${\cal H}^*_{TL}$ to give an effective Hamiltonian in the scaling region
of the new fixed point:
\begin{equation}
{\cal H}_{scaling} = {\cal H}^*_{TL} + \mu {\cal O}(0), \label{scaling}
\end{equation}
with $\mu$ the scaling field conjugate to ${\cal O}$. Unlike for ordinary
critical phenomena, ${\cal O}$ produces the leading scaling in temperature
although it is in fact an irrelevant boundary operator, given by the first
Kac-Moody descendant in the $j=1$ conformal tower: $\vJ_{-1}\cdot \vphi$,
of dimension $\frac{3}{2}$. This is the same operator that drives
critical scaling in the two-channel Kondo effect for noninteracting
electrons \cite{AL}. Specifically, the impurity contributions to
the specific heat $\delta C $ and spin susceptibility $\delta \chi $
are given to leading order by
\begin{equation}
\delta C  =  \frac{\mu^2 9 \pi^2}{v_s^3}
 T \ln (\frac{1}{\tau_0T}),  \ \ \ \ \
\delta \chi  =  \frac{\mu^2 18}{v_s^3} \ln (\frac{1}{\tau_0T}), \ \ \ \ \
\ T \rightarrow 0,
\label{CHIimp}
\end{equation}
with $\tau_0$ a short-time
cutoff. With the known bulk response for the Tomonaga-Luttinger model,
$C = \pi (v^{-1}_c + v^{-1}_s)T/3$ and $\chi = 1/2\pi v_s$
we predict a Wilson ratio
\begin{equation}
R_W = \frac{\delta \chi / \chi}{\delta C / C} =
\frac{4}{3}(1+\frac{v_s}{v_c}).
\label{Wilson}
\end{equation}
For $g \rightarrow 0$ \ ($v_c$, $v_s \rightarrow v_F$), this reduces to
the universal number 8/3 characterizing the ordinary two-channel
Kondo effect \cite{AL}.

Let us now turn to the more realistic case with forward  $(F)$ {\em and}
backward $(B)$ scattering off the impurity. To keep things simple we assume
the  $F$ and $B$ amplitudes to be the same, and hence replace
$H_F$ by
\begin{equation}
{\cal H}_K \equiv {\cal H}_F + {\cal H}_B =
\lambda \sum_{k,l = L, R} \no{ \psidop{k,\sigma}(0)
\mbox{$\frac{1}{2}$} \bsigma{\sigma \mu} \psiop{l,\mu}(0) } \cdot \vS.
\end{equation}
The extra terms, mixing left and right, break the chiral $SU(2)$ {\em and}
chiral $U(1)$ invariance of ${\cal H}^*_{TL}$, and this produces a boundary
operator of dimension $\leq \onehalf$ at the forward scattering fixed point.
Back scattering is thus a relevant perturbation and drives the system
to a new fixed point describing {\em Kondo interaction} in a Luttinger liquid.

With no electron-electron interaction ($g=0$ in (\ref{TL})) we have a free
bulk Hamiltonian ${\cal H}_0$ together with  ${\cal H}_{K}$.
Passing to a basis spanned by definite-parity fields
$\psi_{\pm,\sigma}(x) = [ \psi_{L,\sigma}(x) \pm \psi_{R,\sigma}(-x)]
/\sqrt{2}$, ${\cal H}_0 + {\cal H}_{K}$ transforms into a
two-channel theory, but with the impurity coupled to the electrons
in only one of the channels. This renormalizes to a local Fermi liquid
(like the ordinary 3D Kondo problem), with response
functions scaling analytically with temperature.
However, a different approach must be used for the interacting problem since
${\cal H}^*_{TL}$ is non-local in this basis. Here we exploit the
expectation that {\em any} local impurity interaction, including the Kondo
interaction ${\cal H}_K $, can be substituted by a renormalized boundary
condition on the critical bulk theory \cite{Affleck}.
The equivalent selection rule defines
a fixed point, and by demanding that any associated $LCBO$ must
respect the symmetries of the problem {\em and} correctly reproduce the
non-interacting limit as $g \rightarrow 0$, the possible critical theories
can be deduced. (Note that a selection rule here defines a {\em boundary}
fixed point, and is valid for all values of the marginal bulk
coupling $g$. Hence, given a selection rule,
Fermi liquid scaling must emerge in the limit $g \rightarrow 0$.)

To have a generally applicable formalism we replace $Q$ and $\Delta Q$ by
new quantum numbers, treating the two diagonalized $U(1)$ towers as {\em
independent.} (In the previous case these two towers were tangled
up via the labeling by $Q$ and $\Delta Q$, as an
implicit part of the corresponding selection rule for combining conformal
towers.) This allows us to formulate a general criterion for selecting
operators that respect global, but not necessarily chiral, $U(1)$
invariance. Together with invariance under channel exchange
$(1 \leftrightarrow 2)$, this leaves only certain possibilities for
the $U(1) \times U(1)$ part of the candidate {\em LCBOs}. The complete set of
possible scaling dimensions are then obtained by also including
the $SU(2)_2$ and Ising conformal towers, as allowed by symmetry.

By requiring Fermi-liquid scaling for the impurity specific heat
$\delta C$ and susceptibility $\delta \chi $ as $g \rightarrow 0$,
we find that there are only {\em two} possible types of
critical behavior. {\em Either} Fermi-liquid scaling persists for
$g \neq 0$, {\em or} there is a non-Fermi liquid behavior given by
\begin{eqnarray}
\delta C & = & c_1((1/K_{\rho})-1)^2T^{(1/K_{\rho})-1} + c_2 \, T,
 \label{anomalousC} \\
\delta \chi & = & c_3 \, T^0, \label{anomalousX}
\end{eqnarray}
as $T \rightarrow 0$. Here $K_{\rho} = (1+2g/v_F)^{-1/2}$ and $c_{1,2,3}$
are amplitudes depending on the scaling fields and velocities.
The $LCBO$ driving the anomalous scaling in (\ref{anomalousC}) and
(\ref{anomalousX}) is given by the composite operator
${\cal O}_{LCBO} = [V^1_{2,0} \times V^2_{-2,0} +
V^1_{-2,0} \times V^2_{2,0}] \times \epsilon$ where
$V^i_{C,D}$ is a $U(1)$ primary (vertex) operator in channel $i$,
and $\epsilon$ the Ising energy density.

The scaling in (\ref{anomalousC})
and (\ref{anomalousX}) agrees exactly with a conjecture by
Furusaki and Nagaosa \cite{FN}, in support of the non-Fermi liquid scenario.
However, a simplified model (neglecting backward spin diagonal and forward
spin off-diagonal Kondo scattering) suggests that in fact the other scenario
(Fermi liquid) may be realized \cite{SI}! Whatever is the case, our exact CFT
analysis shows that {\em no other fixed point theories are possible}.
Note that in none of the two cases does the electron-electron
interaction influence $\delta \chi$: the impurity remains completely
screened for $g \neq 0$.

This work is supported by NSF grants DMR-9205125 and DMR-91-120282 (P. F.),
and a grant from the Swedish Natural Science Research Council (H. J.).



\end{document}